\documentclass[a4paper]{jpconf}
\usepackage{amsfonts, amsmath, amsthm, amssymb}
\usepackage{bm}
\usepackage{graphicx}

\begin{document}
\title{Dirac bi-spinor entanglement under local noise and its simulation by Jaynes-Cummings interactions}

\author{Victor A. S. V. Bittencourt and Alex E. Bernardini}

\address{Departamento de F\'{\i}sica, Universidade Federal de S\~ao Carlos, PO Box 676, 13565-905, S\~ao Carlos, SP, Brasil.}

\ead{vbittencourt@df.ufscar.br}

\begin{abstract}
A description of the effects of the local noise on the quantum entanglement constraining the internal degrees of freedom of Dirac bi-spinor structures driven by arbitrary Poincar\'{e} invariant potentials is proposed.  Given that the Dirac equation dynamics including external potentials can be simulated by a suitable four level trapped ion setup, quantum entanglement of two-qubit ionic states with quantum numbers related to the total angular momentum and to its projection onto the direction of the external magnetic field (used for lift the ions degeneracy), are recovered by means of a suitable ansatz. This formalism allows the inclusion of noise effects, which leads to disentanglement in the four level trapped ion quantum system. Our results indicate the role of interactions in bi-spinor entanglement, as well as the  description of disentanglement in ionic states under local noises. For a state prepared initially in one of the ionic levels, local noise induces entanglement sudden death followed by sudden revivals driven by the noiseless dynamics of the state. Residual quantum correlations are observed in the intervals where such state is separable. Schr\"odinger cat and Werner states partially loose their initial entanglement content due to the interaction with the noisy environment but presenting entanglement oscillations without sudden death.  Because Dirac equation describes low energy excitations of mono layer and bi-layer graphene, the formalism can also be applied to compute, for instance, electron-hole or electron/electron entanglement in various circumstances.
\end{abstract}

\section{Introduction}

\hspace{1 em} Quantum correlations has been in the core of recent developments exploring the interface between quantum and classical physics \cite{001, 002, 003, 004}. Quantum entanglement and other quantum correlations are widely considered for the engineering of quantum information protocols, in particular for quantum cryptography \cite{A002,A002B} and quantum computing processes \cite{A004}. The non-local coherence generated by entanglement is essential to various applications of such quantum information/computing tasks in physical systems \cite{A004}.

One platform suitable for implementing and characterizing quantum correlations is the ion-trap technology \cite{Nat01,Nat03}, which has provided a phenomenological access to generate and manipulate quantum correlational properties of the trapped ions \cite{n004,n005,n006,nossopaper}. Moreover, the trapped ion setup has also been engendered for simulation protocols, such as for simulating open quantum systems and quantum phase transitions \cite{Nat02,Nat04,Nat05}. Simulating Dirac equation structures is another example of a quantum system engendered by trapped ion dynamics. Under suitable conditions, the ionic (anti)Jaynes-Cummings ((A)JC) Hamiltonian dynamics can be mapped onto the structure of the Dirac equation, reproducing a series of relativistic-like quantum effects \cite{n001,n002,new01}. On the other hand, the two-{\em qubit} intrinsic entanglement present on the $SU(2) \otimes SU(2)$ bi-spinor structure of Dirac equation solutions is also encoded in the ionic system simulating the Dirac dynamics by quantum numbers related to the total angular momentum and to its projection onto the direction of the external magnetic field (which is applied for lift the degeneracy of the ions internal levels) \cite{nossopaper,n009,n010}.

A quantum system not isolated has its correlational properties affected by the coupling with its environment. Fluctuations of the external magnetic field acts as an environment, causing entanglement degradation and a consequent decoherence \cite{intronoise00, intronoise01, intronoise02}. To include such environment effects on the trapped ion two-{\em qubit} structure considered in \cite{nossopaper}, as well as to elucidate how the local decoherece is related to local and non-local disentanglement, a noise model for the environment can be adopted to describe more realistic setups \cite{Yu01}.

This letter is concerned with the effects of a local noise model, where each subsystem is individually affected by the environment, in the informational content of internal states of a four level trapped ion. The local noise is included via Kraus operators and the dynamics of an arbitrary ionic state is recovered through the connection between the (A)JC dynamics and the Dirac equation structure. Once the Dirac-like spinor density matrix is constructed, the quantum entanglement can be quantified through negativity \cite{n019,QC01}. Due to the coupling with the environment, an initial pure state evolves into a separable mixed state, which can display quantum correlations other than entanglement. In particular, the geometric discord \cite{discord01} shall be adopted as a measure of such quantum correlations.
For quantum states constrained to a specific ionic level, sudden deaths and quantum revivals can be quantified by the intrinsic entanglement. Moreover, those states exhibit residual nonclassical correlations for intervals were the state is separable. On the other hand, when the maximally entangled (Werner and Schr\"odinger cat) quantum states -- prepared as superposition of such internal levels -- are considered, coupling to environment causes a partial loss of their initially entangled properties, but no entanglement sudden death is present. In the three states considered, entanglement oscillations, caused by their noiseless dynamics, are observed.

\section{Trapped ion framework}

The trapped ion framework can be formulated by a Hamiltonian $\hat{H}_{RJC}$ which, in the rotating wave approximation \cite{n012}, includes three interactions, the (anti) Jaynes-Cummings and the carrier interactions. The (A)JC interactions are given by the Hamiltonian
\begin{equation}
\label{eqeA01}
\hat{H}_j ^{(A)JC} = \hbar\, \eta_j \tilde{\Omega}_j \, (\,\hat{\sigma}^{+(-)} a_j e^{+(-)i \phi_{r(b)}} + \hat{\sigma}^{-(+)} a_j^\dagger e^{-(+)i \phi_{r(b)}} \,) + \hbar\, \delta_j \hat{\sigma}_z,
\end{equation}
with $j = x,\,y,\,z$, where the phases $\phi_{r\,(b)}$ describe red(blue)-sideband excitations, $\tilde{\Omega}_j$ are the Rabi frequencies, $\hat{\sigma}^{+ \, (-)}$ are the internal level raising(lowering) operators, $\delta_j$ is an emergent detuning frequency between the external field and the two-level system, and $\eta_j = k \sqrt{\hbar\,/2 \tilde{m} \nu_j}$ is the Lamb-Dicke parameter (where $\tilde{m}$ is the ion mass and $k$ is the wave number of the external field). The JC interaction corresponds to a de-excitation of the internal level  accompanied by an excitation of the vibrational degree of freedom, while the AJC interaction accomplished an excitation of the internal level with an excitation of the vibrational degree of freedom. The carrier interaction is given by
\begin{equation}
\label{eqeA03}
\hat{H}_j^C = \hbar\, \Omega_j (\hat{\sigma}^+ e^{i \phi} + \hat{\sigma}^- e^{-i \phi}),
\end{equation}
which excites the internal levels without changing the vibrational state of the ion. From the experimental perspective, the above described dynamics is observed, for example, in the hyperfine levels ($2 s^2 \, S_{1/2}$) of alkali ions.

From now on, we consider a trapped ion with four internal levels labeled $\{\vert a \rangle, \vert b \rangle, \vert c \rangle, \vert d \rangle \}$. The combination of JC, AJC and carrier interactions acting on such system can be mapped into a Dirac Hamiltonian including tensor and pseudotensor external fields \cite{n001,n002,n003,n004,n005},
\begin{equation}
\label{DiracHam}
\hat{H}_D = \hat{\beta} \, m c^2 + c\, \hat{\bm{\alpha}}\cdot\bm{p} + \hat{\beta} \hat{\bm{\Sigma}} \cdot \left( \kappa \, \bm{\mathcal{E}} \right) + i \hat{\beta} \hat{\bm{\alpha}} \cdot \left( \mu \, \frac{\bm{\mathcal{E}}}{c} \right),
\end{equation}
from which one firstly notices that the Dirac mass and kinetic term reads
\begin{equation}
\label{eqeA04}
\hat{\beta} m c^2 \rightarrow 2 \hbar\, \delta( \hat{\sigma}_z^{ad} + \hat{\sigma}_z ^{bc}),
\end{equation}
\begin{equation}
\label{eqeA05}
c \, \hat{\bm{\alpha}}\cdot \bm{p} \rightarrow 2 \eta \Delta_x \tilde{\Omega}(\hat{\sigma}_x^{ad} + \hat{\sigma}_x^{bc}) p_x + 2 \eta \Delta_y \tilde{\Omega}(\hat{\sigma}_y^{ad} - \hat{\sigma}_y^{bc})p_y + 2 \eta \Delta_z \tilde{\Omega}(\hat{\sigma}_x^{ac} - \hat{\sigma}_x^{bd}) p_z,
\end{equation}
such that, for example, $\hat{\sigma}_z^{ad} + \hat{\sigma}_z ^{bc} \equiv \hat{\beta}$, where the upper indices denote the involved internal levels \cite{n003,n004,n005,nossopaper}. The momenta $p_j$ are given by
$p_j \rightarrow \frac{i \hbar\,}{2 \Delta_j} \, (\,a_j ^\dagger - a_j \,)$,
where $\Delta_j = \sqrt{\hbar\,/ 2 \tilde{m} \nu_j}$ is the delocalization width of the ground state wave function.

The tensor and pseudotensor external potentials are mapped in terms of the carrier interactions (\ref{eqeA03}) with frequencies $\Omega_j^{(1)}$ and $\Omega_j^{(2)}$:
\begin{subequations}
\label{eqeA06}
\begin{equation}
\hat{\beta} \hat{\bm{\Sigma}} \cdot \left( \kappa \, \bm{\mathcal{E}} \right) \rightarrow 2 \hbar\, \Omega_x ^{(1)} \, (\, \hat{\sigma}_x^{ab} - \hat{\sigma}_x^{cd}\,) \, + \, 2 \hbar\, \Omega_y ^{(1)} \,(\,\hat{\sigma}_y^{ab} - \hat{\sigma}_y^{cd} \,)\, + \, 2 \hbar\, \Omega_z ^{(1)}\, (\, \hat{\sigma}_z^{ab} - \hat{\sigma}_z^{cd} \,),
\end{equation}
\begin{equation}
i \hat{\beta} \hat{\bm{\alpha}} \cdot \left( \mu \, \frac{\bm{\mathcal{E}}}{c} \right) \rightarrow 2 \hbar\, \Omega_x ^{(2)}\, (\, -\hat{\sigma}_y^{ad} - \hat{\sigma}_y^{bc}\,) \, + \, 2 \hbar\, \Omega_y ^{(2)}\,(\, \hat{\sigma}_x^{bc} - \hat{\sigma}_x^{ad}\,) \, + \,2 \hbar\, \Omega_z ^{(2)} \,(\, \hat{\sigma}_y^{bd} - \hat{\sigma}_y^{ac} \,).
\end{equation}
\label{eqeA06B}
\end{subequations}

The Dirac dynamics driven by (\ref{DiracHam}) is reproduced in the trapped ion setup through the maps (\ref{eqeA04}), (\ref{eqeA05}) and (\ref{eqeA06B}) and the relations between Dirac and ionic parameters are identified as
\begin{equation}
\frac{\mu\, \mathcal{E}_j}{c} = 2 \hbar\, \Omega_j ^{(2)}, \hspace{1.3 cm} \kappa\, \mathcal{E}_j = 2 \hbar\, \Omega_j ^{(1)},\hspace{1.3 cm}
c = 2 \eta \Delta \tilde{\Omega}, \hspace{1.3 cm} m c^2 = 2 \hbar\, \delta,
\end{equation}
which set a one-to-one correspondence between (\ref{DiracHam}) and the sum of the interactions (\ref{eqeA04}), (\ref{eqeA05}) and (\ref{eqeA06B}). Accordingly, the bi-spinor eigenstates of (\ref{DiracHam}) $\vert \psi_{n,s} \rangle$ (with $n,\,s = 0,\,1$), are described as linear combinations of the internal ionic levels,
\begin{equation}
\label{eqeA07}
\vert \psi_{n,s} \rangle \rightarrow M^a_{n,s} \vert a \rangle + M^b_{n,s} \vert b \rangle + M^c_{n,s} \vert c \rangle + M^d_{n,s} \vert d \rangle.
\end{equation}
Each of the ionic state $\vert i \rangle$ is characterized by two quantum number: the total angular momentum $F$, and the projection $M$ of the angular momentum onto an external magnetic field used for lift the degeneracy of the internal ionic levels. Such structure allows the assignment of a two-{\em qubit} correspondence given by
\begin{equation}
\label{eqeA08}
\vert a \rangle \equiv\vert 0 \, 0 \rangle, \hspace{1.5 cm} \vert b \rangle \equiv \vert 0 \, 1 \rangle, \hspace{1.5 cm}
\vert c \rangle \equiv \vert 1 \, 0 \rangle, \hspace{1.5 cm} \vert d \rangle \equiv \vert 1 \, 1 \rangle.
\end{equation}

According to the analysis of the classes of Poincar\'e invariant Dirac-like interactions \cite{n009,nossopaper} the Hamiltonian eigenstates $\vert \psi_{n,s} \rangle$ indeed exhibit a naturally \textit{spin-parity} entangled structure \cite{salomon} which can be straightforwardly computed from stationary pure states, $\varrho_{n,s} = \vert \psi_{n,s} \rangle \langle \psi_{n,s} \vert$, given by \cite{n009}
\begin{eqnarray}
\label{varho}
\varrho_{n,s} &=& \frac{1}{4} \left[ I +\frac{ (-1)^n}{\vert \lambda_{n,s} \vert} \hat{H}_D \right] \left[I + \frac{(-1)^s}{\sqrt{g_2}} \hat{\mathcal{O}} \right], \nonumber \\
\hat{\mathcal{O}} &=& \frac{1}{2}\left(\hat{H}_D ^2 - \hat{I} \right) = m\,\kappa\, \hat{\bm{\Sigma}} \cdot \bm{\mathcal{E}} + \mu \, \hat{\beta}\,\hat{\bm{\Sigma}} \cdot \, (\, \bm{p} \times \bm{\mathcal{E}} \,) - i \kappa\,\hat{\beta} \,\hat{\bm{\alpha}} \cdot (\, \bm{p} \times \bm{\mathcal{E}} \,),
\end{eqnarray}
where the parameter $g_2$ is evaluated as $g_2 = \frac{1}{16}\Tr\left[\left(\hat{H}_D^2- \frac{1}{4}\Tr[\hat{H}_D^{2}]\right)^{2}\right]$ and $\lambda_{n,s}$ is identified as the mean energy of $\varrho_{n,s}$, $\lambda_{n,s} = \mbox{Tr} [\, \hat{H}_D \, \varrho_{n,s}\,]$.
Considering a Dirac-like one-dimensional propagation along the $x$ axis, with the electric field lying in the $xy$-plane, such that
$\bm{\mathcal{E}} = \mathcal{E}(\, \cos{\theta}\,\bm{i} + \sin{\theta} \,\bm{j} \,) \nonumber $, with
$\bm{p} = p \,\bm{i}$, where $\bm{i},\bm{j},\bm{k}$ define an orthonormal basis, one has $\bm{p}\times \bm{\mathcal{E}} = p \, \mathcal{E} \sin{\theta} \, \bm{k}$.
The arbitrary choice of $\theta$ does not qualitatively affect the results, thus, one shall adopt $\theta = \pi/4$, such that the corresponding expressions for $g_2$ and $\lambda_{n,s}$ read
\begin{subequations}
\begin{eqnarray}
g_2 &=& \mathcal{E}^2 \left[\, m^2 \kappa^2 + \frac{1}{2}(\mu^2 + \kappa^2) p^2 \, \right], \\
\lambda_{n,s} &=& (-1)^n\bigg[\, p^2 + m^2 + (\kappa^2 + \mu^2)\mathcal{E}^2 + 2 (-1)^s \mathcal{E}\sqrt{m^2 \kappa^2 + \frac{1}{2}(\mu^2 + \kappa^2)\,p^2 \,} \bigg]^ {1/2}.
\end{eqnarray}
\end{subequations}

To recover the dynamics of an internal ionic state, the Dirac bi-spinor basis, $\{ \vert \, \psi_{n,s} \, \rangle \}$, shall be related to the ionic state basis, $\{ \vert\, i \, \rangle \}$. The temporal evolution of a single internal level is thus reconstructed by using the completeness relation $\displaystyle \sum^1_{n,s = 0} \varrho_{n,s} = \hat{I}$ as to have (for $\vert j (t=0) \rangle \equiv \vert j \rangle$)
\begin{equation}
\rho_j (t) = \vert\, j (t)\, \rangle \langle\, j(t) \, \vert
= e^{- i \hat{H}_D t} \vert j \rangle \langle j \vert e^{ i \hat{H}_D t} = \displaystyle \sum^1_{n,s=0}\sum^1_{m,l=0} e^{- i (\lambda_{n,s} - \lambda_{m,l}) t}\, \varrho_{n,s} \, \rho_j (0) \, \varrho_{m,l}.
\label{internalionic}
\end{equation}
A typical four level quantum oscillation pattern is present in this time evolution. For instance, a state initially in $\vert j \rangle$ evolves into a generic superposition, with conversion probability to a particular ionic level $\vert k \rangle \neq \vert j \rangle$ obtained through (\ref{internalionic}) \cite{nossopaper}.

The assignment (\ref{eqeA08}) suggests the identification of two subsystems - $\mathcal{S}_F$, related to the total angular momentum quantum number $F$, and $\mathcal{S}_{M}$ associated to the projection of the angular momentum onto the direction of the projection $M$ of the external magnetic field. In this framework, a state initially prepared in the internal state $\vert j \rangle$ will evolve to a superposition between the four internal states and shall exhibit quantum entanglement (and possibly other quantum correlations in the case of mixed states) between the two subsystems $\mathcal{S}_F$ and $\mathcal{S}_{M}$.

To describe more realistic setups, one is inclined to consider environment effects on the trapped ion dynamics. The inclusion of such open systems effects can be accomplished by an effective Hamiltonian $\hat{H}_{Env}$ describing environment coupling as to have a total Hamiltonian given by
\begin{equation}
\label{TotHamil}
\hat{H} = \hat{H}_j ^{(A)JC} + \hat{H}_{Env}.
\end{equation}
In particular, the influence of an environment is included through the influence of two \textit{local} noises, generated by random fluctuations acting separately on each qubit which is given by the Hamiltonian \cite{Yu01}:
\begin{eqnarray}
\hat{H}^{\rm Loc}_{\rm Env} &=&
- \frac{1}{2} \mu \left(\, b_1 (t) \,( \hat{\sigma}^{(1)}_z \otimes \hat{I} ^{(2)}) +  b_2(t) \,(\hat{I} ^{(1)} \otimes \hat{\sigma}_z ^{(2)}) \,\right).
\label{eqe02}
\end{eqnarray}
In this model, separate fluctuations act on the corresponding subsystem and the fields $b_i (t)$ are characterized by Markovian conditions
\begin{eqnarray}
\langle b_i (t) \rangle = 0 ,  \hspace{0.5 cm} \langle b_i (t) b_i (t^\prime) \rangle = \frac{\Gamma_i}{\mu^2} \delta(t - t^\prime),
\end{eqnarray}
were $\langle \dots \rangle$ represent the ensemble average and $\Gamma_i$ are the phase relaxations due to the local interactions. This setup was previously considered for study phonon decoherence \cite{Yu02}, for protocols to measure magnetic field gradient \cite{magnetic} in the trapped ion framework, and for investigating the influence of noise on transport properties in chains of trapped ions \cite{noisetransport}. Here it is supposed that the fluctuations on the degeneracy lifting magnetic field affect each degree of freedom individually. Other often considered noise model is the \textit{global} noise \cite{Yu01} in which the random fluctuations act equally on both qubits. The influence of such environment will be investigated in a future issue.

\section{Dynamics and Results}

The temporal evolution driven by the total Hamiltonian $\hat{H}$ (\ref{TotHamil}), can be recovered by decoupling it into two pieces through the interaction picture:
\begin{equation}
\tilde{\rho}(t) = e^{i\int^t_0 \hat{H}_{\rm Env}(s)ds} \rho(0)
e^{-i\int^t_0 \hat{H}_{\rm Env}ds},
\label{master01}
\end{equation}
with $\rho(0)$ being the initial state and $\tilde{\rho}(t)$ describing the time-evolved density matrix in the interaction picture, which is a solution of the corresponding master equation given in terms of the Kraus representation \cite{A004,kra,wk}. The inclusion of noise effects through Kraus representation is thus implemented by taking the statistical mean of Eq.~(\ref{master01}) \cite{Yu01}.
The behavior of $\tilde{\rho}(t)$ under the influence of the local noise described by the Hamiltonian (\ref{eqe02}) is then expressed in a compact way as
\begin{equation}
\label{globalsoln}
\tilde{\rho}(t)={\mathcal E}_D(\rho(0)) = \sum_{\mu, \nu = 1} ^{2} E_\mu^\dag(t) F_\nu ^\dagger (t)
\rho(0) F_\nu(t) E_\mu (t),
\end{equation}
where the Kraus operators describing the local interaction (\ref{eqe02}) are given by:
\begin{eqnarray}
E_1 &=& \mbox{diag}\{1, \gamma_1\} \otimes \hat{I}^{(2)}, ~E_2 = \mbox{diag}\{0, \omega_1\} \otimes \hat{I}^{(2)}, \nonumber \\
F_1 &=& \hat{I}^{(1)} \otimes \mbox{diag}\{1, \gamma_2\},~ F_2 = \hat{I}^{(1)} \otimes \mbox{diag}\{0, \omega_2\},
\end{eqnarray}
with $\gamma_i = e^{-\Gamma_i t/2}$ and $\omega_i = \sqrt{1-e^{- \Gamma_i \,t}}$. For simplicity we adopted equal phase relaxations on both subsystems, i.e. $\Gamma_1 = \Gamma_2 = \Gamma$.

After the above described procedure, to recover Schr\"odinger picture density matrix one uses the completeness relation for the eigenstates $\{ \varrho_{n,s} \}$ as to have
\begin{eqnarray}
\label{eqeD01}
\rho(t) = e^{i \hat{H}_D t} \tilde{\rho}(t) e^{ - i \hat{H}_D t}
= \displaystyle \sum^1_{n,s = 0} \sum^1_{m,l = 0} \sum_{\mu=1}^3 e^{ - i (\lambda_{n,s} - \lambda_{m,l}) t} \, \varrho_{n,s} \,D_\mu^\dag(t)\, \rho(0)\, D_\mu(t) \, \varrho_{m,l},
\end{eqnarray}
from which any observable can be evaluated for any given initial state $\rho(0)$.

From now on, the disentangling properties of ionic states under local noise effects can be investigated. In particular we shall consider the evolution of the internal ionic state $\vert a (0) \rangle = \vert a \rangle$, and the evolution of the Schr\"odinger cat state $\rho_C(0) = \vert \psi_C \rangle \langle \psi_C \vert$ and of the Werner state $\rho_W (0) = \vert \psi_W \rangle \langle \psi_W \vert$, which are written in terms of internal ionic states as
$\vert \psi_C \rangle = \frac{\vert a \rangle + \vert d \rangle}{\sqrt{2}}$ and
$\vert \psi_W \rangle = \frac{\vert b \rangle + \vert c \rangle}{\sqrt{2}}$.
The time evolutions of these states are recovered by Eq.~(\ref{eqeD01}) from which all correlational properties can be computed. Accordingly to the Peres criterion, any separable state must have all eigenvalues of its partial transpose density matrix positive \cite{n019}. The negativity of a state $\varrho$ is thus defined by \cite{QC01}
\begin{equation}
\mathcal{N}[\varrho] = \vert \vert \, \varrho^T_A \, \vert \vert - 1 = \displaystyle \sum_i \vert \mu_i \vert -1,
\end{equation}
where $\vert \vert \, \varrho^T_1 \, \vert \vert = \sum_i \vert \mu_i \vert$ stands for the trace norm of the partial transposed matrix $\varrho^T_1$, whose eigenvalues are $\{ \mu_i \}$. Roughly speaking, negativity measures the extent to which the partial transpose fails to be positive.

Separable mixed states can exhibit quantum correlations other than entanglement \cite{QC02}, and the characterization of such quantum correlations is still an open problem. The quantum discord, defined as the difference between total correlations among two subsystems before and after the action of perfect local measurements in one of them, is a quantifier for nonclassical correlations \cite{QC03}. The evaluation of quantum discord requires an extremization procedure over all possible sets of local measurements in one of the subsystems, being a NP-hard problem with no analytical solution or even efficient algorithm (unless $P= NP$) \cite{QUANTUMDISC}. To avoid such computational issue, we shall adopt the geometric discord $D[\varrho]$, as a measure for quantum correlations \cite{QC04}. Such quantity is defined as the minimum distance between the state $\rho$ and the set of states with zero quantum discord. Given the Fano decomposition of a state $\varrho$ $$\varrho = \frac{1}{4} \left[\hat{I}+(\hat{\bm{\sigma}}^{(1)} \otimes \hat{I}^{(2)}) \cdot \bm{a}^{(1)}+(\hat{I}^{(1)} \otimes \hat{\bm{\sigma}}^{(2)}) \cdot \bm{a}^{(2)} + \displaystyle \sum_{i,j = 1}^3 t_{ij} \hat{\sigma}_i ^{(1)} \otimes \hat{\sigma}_j ^{(2)}\right],$$ geometric discord associated with the $i$ subsystem of the state is analytically computed as \cite{discord01}
\begin{equation}
D[\varrho]_{1 (2)} = \frac{1}{4} \left(a_{1 (2)}^2 + \parallel T \parallel^2 - k_{max} \right),
\end{equation}
where $\parallel T \parallel^2 = Tr[T T^T]$, $T$ being the matrix composed by the $t_{ij}$ coefficients, and $k_{max}$ is the largest eigenvalue of $\bm{a}_{1(2)} \bm{a}_{1 (2)}^T +T T^T$. It was proved that $D[\varrho]=0$ is a necessary and sufficient condition for vanishing quantum discord \cite{discord01}, and that the geometric discord and negativity satisfy the inequality $(\mathcal{N}[\varrho])^2 \le D[\varrho]$ \cite{QC05}.

\begin{figure}
\includegraphics[width = 7.5 cm]{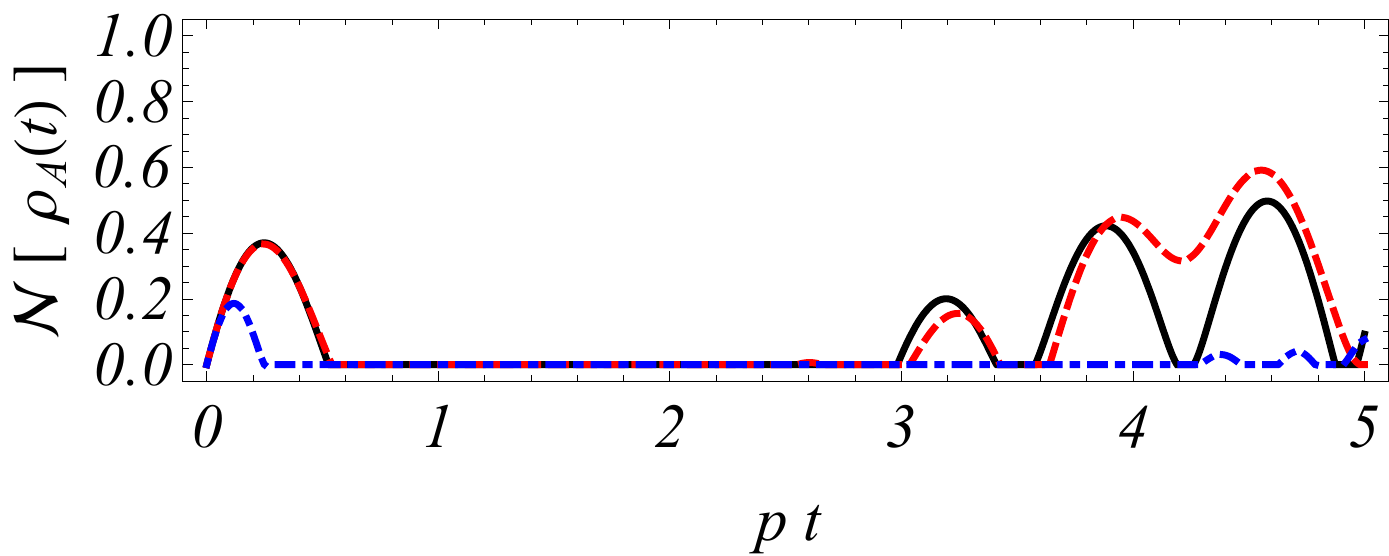}
\includegraphics[width = 7.5 cm]{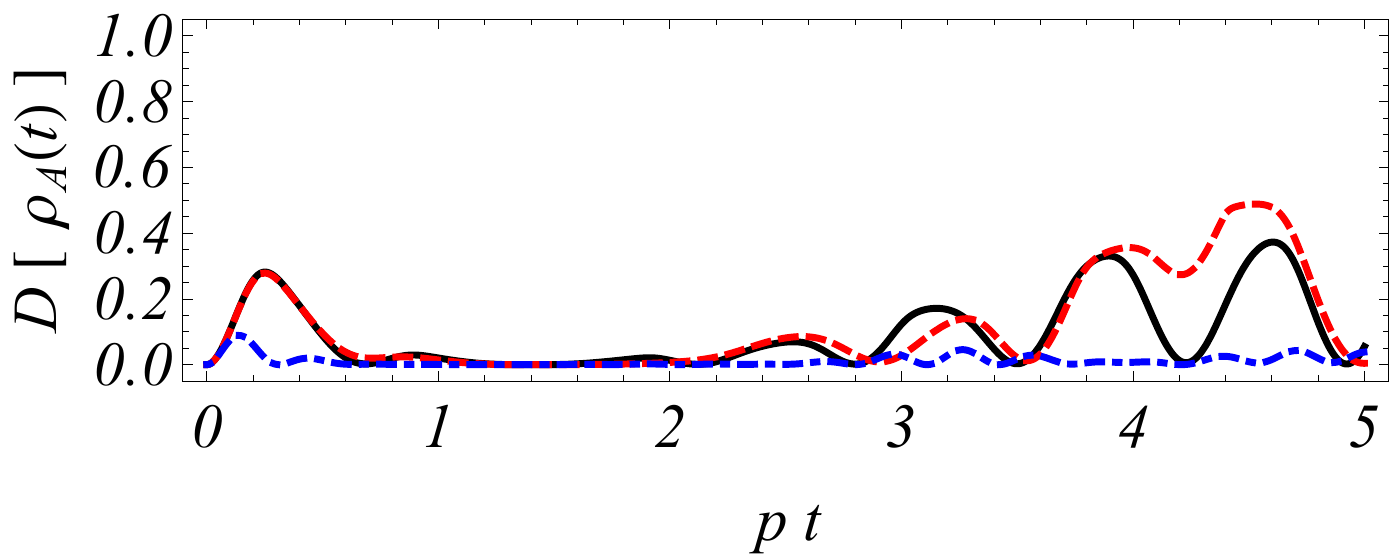}
\caption{Entanglement (left plot) and quantum correlations (right plot) of the state $\vert a (t) \rangle$ as function of $p \, t$ for $\kappa = \mu = 1$, $\mathcal{E}/p = 1$, $\Gamma/p = 1/2$ and for $m/p = 0$ (solid curves), $1$ (dashed curves), 10 (dot-dashed curves). The influence of the local noise causes entanglement sudden death, followed by sudden revivals and quantum oscillations generated by the noiseless evolution of the state. Moreover, in the periods where the mixed state is separable, only residual quantum correlations are present, as depicted in the right plot.}
\label{eqefig:01}
\end{figure}

The correlational properties of the state $\rho_A (t) = \vert a(t) \rangle \langle a(t) \vert$ are depicted in Fig.~\ref{eqefig:01} which shows the negativity (left plot) and the geometric discord (right plot) as function of $p \, t$ for $\kappa = \mu = 1$, $\mathcal{E}/p = 1$, $\Gamma/p = 1/2$ and for $m/p = 0$ (solid curves), $1$ (dashed curves), 10 (dot-dashed curves). The noiseless evolution drives the state to a superposition among all the internal states for $t>0$, generating intrinsic entanglement. Nevertheless the influence of the local noise suddenly makes the entanglement vanish in an effect known as entanglement sudden death \cite{Diosi}, which is followed by sudden revivals in a oscillation pattern without definite frequency. In the intervals where the state is separable, only residual quantum correlations persists, as depicted in the left plot of Fig.~\ref{eqefig:01} and, similar to the negativity, geometric discord oscillates without a definite frequency. Moreover, large values of $m/p$ corresponds to a stronger influence of the environment on the correlations.

\begin{figure}
\includegraphics[width = 7.5 cm]{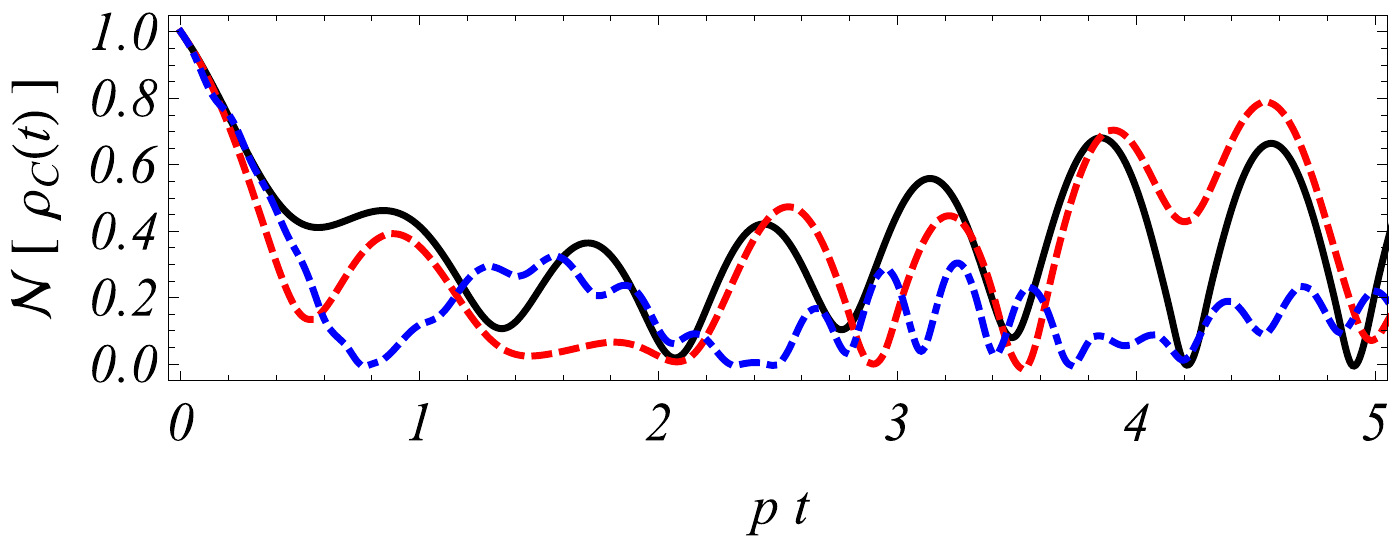}
\includegraphics[width = 7.5 cm]{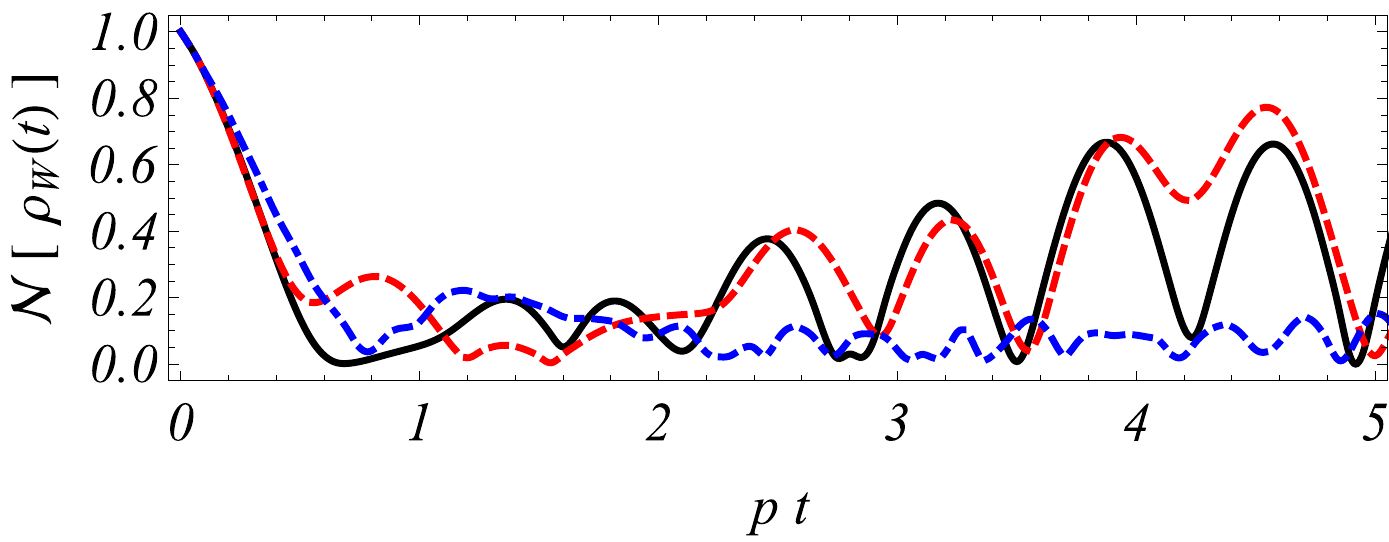}
\caption{Werner (right plot) and cat (left plot) state entanglement under local noise as function of $p\,t$. The parameters and plot styles are in correspondence with those of Fig.~\ref{eqefig:01}. The local noise dissipates the entanglement in both states, but for low values of $m/p$ entanglement oscillations are still present. For $t \gg 1/\Gamma$, both Werner and cat states are pure, being mixed only for intermediate values of $p\,t$.
}
\label{eqefig:02}
\end{figure}

Under local noise, both Werner and Schr\"{o}dinger cat states suffer decoherence, as depicted in Fig.~\ref{eqefig:02}. Entanglement is partially dissipated, nevertheless, different from the previous case, no entanglement sudden death is observed and negativity oscillates without a definite periodicity. Even for $t \gg 1/\Gamma$, the states exhibit entanglement. Intermediate values of $m/p$ characterize a stronger influence of the local noise on the entangling properties of such states, as depicted in the blue curves of Fig.~\ref{eqefig:02}, which correspond to $m/p = 10$. The Werner and Schr\"odinger cat states are, qualitatively, equally affected by the local noise, and both are pure states for $t \gg 1/\Gamma$.

\section{Conclusions}

The issue investigated throughout this letter is concerned with the inclusion of local noise effects, described by stochastic fields acting on four ionic levels, in the context of a Dirac-like dynamics simulated by the trapped ion systems.
The entanglement properties were quantified in terms of negativity and the quantum correlations were quantified by means of the geometric discord, where the noiseless dynamics of ionic states was recovered through the correspondence between JC, AJC and carrier interactions through the inclusion of tensor and pseudotensor Dirac structures into the simulated Hamiltonian.
The noise was introduced by means of the Kraus representation formalism, such that quantum correlations of states prepared as internal levels of the trapped ion were then investigated.
The quantum correlational properties exhibited by a state initially on a single level, and by Werner and Schr\"odinger cat states formed by superpositions between two internal ionic levels were finally obtained.

Under local noise, for Werner and Schr\"odinger cat states, the entanglement is partially lost, but entanglement oscillations are still present due to the pureness of both states for $t \gg 1/\Gamma$. The mass parameter on the Dirac Hamiltonian actively suppresses the entanglement content of the states. For a state initially in the internal level $a$, entanglement shows sudden deaths and sudden revivals, which are driven by the coupled dynamics with the environment. For such state, the noise also suppress quantum correlations other than entanglement, and in the intervals where the state is separable it exhibits residual quantum correlations.

One concludes by noticing that such trapped ion platforms can work to implement more complex quantum simulations \cite{Gerritsma} where, from the beginning, the only observable that can be  measured (through fluorescence techniques) still is $\hat{\sigma}_z$, related to issues that deserve some further investigations.

{\ack Acknowledgments - The work of AEB was supported by the Brazilian Agency FAPESP (grant 15/05903-4). The work of VASVB was supported by the Brazilian Agency CNPq (grant 140900/2014-4).}


\section*{References}

\begin{thebibliography}{99}
\bibitem{001}
Zurek W H 1991 \textit{Phys. Today} {\bf 44} (10) 36; 1981 {\em Phys. Rev.} {\bf D 24} 1516; 1982 {\em Phys. Rev.} {\bf D 26} 1862

\bibitem{002}
Joos E and Zeh H D 1985 \textit{Z. Phys.} {\bf B 59} 223

\bibitem{003}
Braun D, Haake F and Strunz W T 2001 \textit{Phys. Rev. Lett.} {\bf 86} 2913

\bibitem{004}
Anastopoulos C and Hu B L 2000 \textit{Phys. Rev.} {\bf A 62} 033821

\bibitem{A002}
Shore B W 1990 {\em The Theory of Coherent Atomic Excitation}, vols. 1 and 2 (New York: Wiley), Chap. 23

\bibitem{A002B}
Gisin N, Ribordy G, Tittel W and Zbinden W 2002 {\em Rev. Mod. Phys.} {\bf 74} 145

\bibitem{A004}
Nielsen M A and Chuang I L 2000 {\em Quantum Computation and Quantum Information} (England: Cambridge University Press)

\bibitem{Nat01}
Gessner M, Ramm M, Pruttivarasin T, Buchleitner A, Breuer H P and Haeffner H 2014 {\em Nature Physics} {\bf 10} 105

\bibitem{Nat03}
Ospelkaus C, Warring U, Colombe Y, Brown K R, Amini J M, Leibfried D and Wineland D J 2011 {\em Nature} {\bf 476} 181

\bibitem{n004}
Bermudez A, Martin-Delgado M A and Solano E 2007 \textit{Phys. Rev.} {\bf A 76}, 041801(R) 

\bibitem{n005}
Tenev T G, Ivanov P A and Vitanov N V 2013 {\em Phys. Rev.} {\bf A 87} 022103 

\bibitem{n006}
Lamata L, Casanova J, Gerritsma R, Roos C F, Garc\'ia-Ripoll  J J and Solano E 2011 {\em New Journal of Physics} {\bf 13} 095003 

\bibitem{nossopaper}
Bittencourt V A S V, Bernardini A E, Blasone M 2016 {\em Phys. Rev.} {\bf A 93} 053823 

\bibitem{Nat02}
Schindler P, M\"uller M, Nigg D, Barreiro J T, Martinez  E A, Hennrich M, Monz T, Diehl S, Zoller P and Blatt R 2013 {\em Nature Physics} {\bf 9} 361

\bibitem{Nat04}
Barreiro J T, M\"uller M, Schindler P, Nigg D, Monz T, Chwalla M, Hennrich M, Roos C F, Zoller P and Blatt R 2011 {\em Nature} {\bf 470} 486

\bibitem{Nat05}
Islam R {\em  et al} 2011 {\em Nature Commun.} {\bf 2} 377

\bibitem{n001}
Lamata L, Le\'{o}n J, Sch\"{a}tz T and Solano E 2007 {\em Phys. Rev. Lett.} {\bf 98} 253005 

\bibitem{n002}
Casanova J, Garc\'{\i}a-Ripoll J J, Gerritsma R, Roos C F and Solano E 2010 {\em Phys. Rev.} {\bf A 82} 020101(R) 

\bibitem{new01}
Lee T E, Alvarez-Rodriguez U, Cheng X H, Lamata L and Solano E 2015 {\em Phys. Rev.} {\bf A 92} 032129 

\bibitem{n009}
Bittencourt V A S V and Bernardini A E 2016 {\em Annals of Physics} {\bf 364} 182 

\bibitem{n010}
Bittencourt V A S V, Mizrahi S S and Bernardini A E 2015 {\em Annals of Physics} {\bf 355} 35

\bibitem{intronoise00}
Leibfried D, Blatt R, Monroe C and Wineland D 2003 {\em Rev. Mod. Phys.} {\bf 75} 281

\bibitem{intronoise01}
Kielpinski D, Monroe C and Wineland D J 2002 {\em Nature} {\bf 417} 709 

\bibitem{intronoise02}
Wineland D J, Monroe  C, Itano W M, Leibfried D, King B E and Meekhof D M 1998 {\em J. Res. Natl. Inst. Stand. Technol.} {\bf 103} 259 

\bibitem{Yu01}
Yu T and Eberly J H 2006 {\em Opt. Commun.} {\bf 264} 393; 2004 {\em Phys. Rev. Lett.} {\bf 93} 140404; 2003 {\em Phys. Rev.} {\bf B 68} 165322

\bibitem{n019}
Peres A 1995 {\em Phys. Rev. Lett.} {\bf 77} 8 

\bibitem{QC01}
Vidal G and Werner R F 2002 {\em Phys. Rev.} {\bf A 65} 032314 

\bibitem{discord01}
Daki\'{c} B, Vedral V and Brukner C 2010 {\em Phys. Rev. Lett.} {\bf 105} 190502 

\bibitem{n012}
Liebfried D, Blatt R, Monroe C and Wineland D 2003 {\em Rev. Mod. Phys.} {\bf 75} 281 

\bibitem{n003}
Gerritsma R, Lanyon B P, Kirchmair G, Z\"ahringer F, Hempel C, Casanova J, Garc\'ia-Ripoll J J, Solano E, Blatt R and Roos C F 2011 {\em Phys. Rev. Lett.} {\bf 106} 060503 

\bibitem{salomon}
Bernardini A E and Mizrahi S S 2014 {\em Physica Scripta} {\bf 89} 075105 

\bibitem{Yu02}
Yu T and Eberly J H 2002 {\em Phys. Rev.} {\bf B 66} 193306 

\bibitem{magnetic}
Ng H T and Kim K 2014 {\em Optics Communication} {\bf 331} 353-358 

\bibitem{noisetransport}
Cormick C and Schmiegelow C T 2016 {\em Phys. Rev.} {\bf A 94} 053406 

\bibitem{kra}
Kraus K 1983 {\em States, Effects, and Operations: Fundamental Notions in Quantum Theory} (Berlin: Springer-Verlag)

\bibitem{wk}
W\'odkiewicz K 2001 {\em Opt. Express} {\bf 8} 145

\bibitem{QC02}
Henderson L and Vedral V 2001 {\em J. Phys. A: Math. Gen.} {\bf 34} 6899-6905 

\bibitem{QC03}
Ollivier H and Zurek W H 2001 {\em Phys. Rev. Lett.} {\bf 88} 017901 

\bibitem{QUANTUMDISC}
Huang Y 2014 {\em New Journal of Physics} {\bf 16} 033027 

\bibitem{QC04}
Daki\'{c} B, Vedral V and Brukner C 2010 {\em Phys. Rev. Lett.} {\bf 105} 190502 

\bibitem{QC05}
Girolami D and Adesso G 2011 {\em Phys. Rev.} {\bf A 84} 052110 

\bibitem{Diosi}
Di\'{o}si L 2003 {\em Lect. Notes Phys.} {\bf 622} 157-163 

\bibitem{Gerritsma}
Gerritsma R, Kirchmais G, Zahringer F, Solando E, Blatt R and Ross C F 2010 {\em Nature} {\bf 463} 68

\end{thebibliography}
\end{document}